\documentclass{phbauth}
\usepackage{graphicx}

\begin{document}

\begin{frontmatter}


\title{Pulsed-laser beam profiling for deposition
of high-T$_\mathrm{c}$ YBCO films}

\author{Evgeny V.\,Pechen\thanksref{thank1}}

\address{Lebedev's Physics Institute, RAS,
Leninskij Prosp.53, 117924 Moscow, Russia}

\thanks[thank1]{E-mail: pechen@sci.lpi.msk.su}

\begin{abstract} 

A special shaping of YAG:Nd$^\mathrm{3+}$ laser beam profile has been found
to improve homogeneity and $T_\mathrm{c}$ of
YBa$_\mathrm{2}$Cu$_\mathrm{3}$O$_\mathrm{7-\delta}$ 
films being deposited by the pulsed laser.  
Targets of a considerably non-stoichiometric composition and substrates of 
slanting cuts have turned to be necessary for the high-quality film growing.
Multiple droplets and solid particles ejected from the
target surface by the infrared radiation are removed successfuly
from the laser-induced plasma by use of a velocity filter. 

\end{abstract}


\end{frontmatter}

\section{Introduction}
Conventional on-axis (as opposed to off-axis\,\cite{off-axis}) pulsed
laser deposition is one of the most convenient methods for growing 
films of high-temperature superconductors
(HTS)\,\cite{{on-axis1},{on-axis2}} and HTS-based multilayer 
structures\,\cite{on-axis3}.

A large number of droplets and other macro-particles arriving at
the substrates is the main problem of the method. Use of
a fast shutter for velocity filtration of plasma streams induced by
a short-wave excimer laser radiation (wavelength 308~nm or shorter)
has been found to be an effective method for protecting substrates
from hit of the particles\,\cite{APL}. 

\indent
As compared to the excimer lasers  
YAG:Nd$^\mathrm{3+}$ lasers are handy and cheaper in use however
the radiation density across their beams usually is non-uniform.
Furthermore, they produce radiation of longer (1.06 $\mu$m) wavelength
and generate much more droplets and solid particles of larger size,
therefore.

\indent
In this paper new conditions of on-axis preparation of 
YBa$_\mathrm{2}$Cu$_\mathrm{3}$O$_\mathrm{7-\delta}$ (YBCO) films
by YAG:Nd$^\mathrm{3+}$ pulsed laser are studied. 

\section{Experiment}
\indent
The YBCO film deposition was performed in oxygen atmosphere at a pressure
of 0.3 mbar on ZrO$_\mathrm{2}$:Y$_\mathrm{2}$O$_\mathrm{3}$ (YSZ)
and SrTiO$_\mathrm{3}$ single-crystal
substrates heated to a temperature of 680$^\mathrm{o}$\,C 
and placed at a distance 5.5~cm from a rotated
Y-Ba-Cu-O ceramic target. A disc-chopper installed between the target
and the substrate at a distance 3.8~cm from the target was used as a
fast shutter. An opening of 2.5~cm diameter was
milled out 6 cm off the disk center. The YAG:Nd$^\mathrm{3+}$ laser
pulses of 10~ns duration, 0.1-0.4~J energy, and about 14 Hz
repetition rate were triggered by a phase-adjustable 
electronic device when the opening situated opposite
the substrate. $T_\mathrm{c}$ of the films was measured as the
temperature of the alternating-magnetic-field half-screening (i.e. when
the susceptibility $\chi'=-0.5$).   

\section{Results}
\indent
Starting the on-axis deposition with targets of 
YBa$_\mathrm{2}$Cu$_\mathrm{3}$O$_\mathrm{7-\delta}$ 
composition we got non-stoichiometric films with low $T_\mathrm{c}$ and
a large amount of precipitates on the surfaces. Changing the target
composition by small 
steps we tested the properties of the films deposited in various conditions. 
The highest $T_\mathrm{c}$ and no precipitates had the films of nearly
stoichiometric (or with a small Ba deficit) composition, however we found
that their preparation required the targets of about
YBa$_\mathrm{1.5}$Cu$_\mathrm{2}$O$_\mathrm{x}$ composition.

\indent 
The ordinary on-axis deposition of 200\,nm thick films 
using the infrared exciting radiation led to extremely large number 
of droplets (up to 10$^8$ cm$^{-2}$). 
As the size of the particles was typically of 1\,$\mu$m and reached 
5-10\,$\mu$m, they  covered the film surface almost completely and
caused a dramatic $T_\mathrm{c}$ reduction. The disc-chopper rotated with
a revolution rate higher than 200 Hz demonstrated an excellent separation
of the fast vapor stream and the slow macro-particles by time of flight,
the $T_\mathrm{c}$ increasing (see Fig.\,1) and the surface becoming
smooth. Practically complete detaining the particles (reduction by a
factor of 10$^\mathrm{6}$, at least) was observed with the rotation
faster than 300 Hz.
\begin{figure}[ht]
\begin{center}\leavevmode
\includegraphics[width=1.0\linewidth]{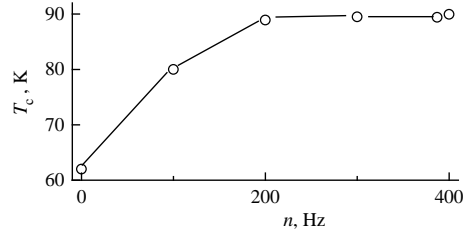}
\caption{ 
$T_\mathrm{c}$ of the YBCO films grown on YSZ substrates with 
different rotation frequency $n$ of the disc-chopper.  
}
\label{fig1}
\end{center}
\end{figure}

\indent
We revealed also a dramatic $T_\mathrm{c}$ reduction with increasing
non-uniformity of the radiation density. So far as highly uniform beam
profile was difficult to achieve and maintain, we increased the mean pulse
energy density from 5-8 to 10-20 J/cm$^2$ so reducing the
vapor composition fluctuations due to complete evaporation
of the less-irradiated target zones. However, it led to a ball-like
plasma shape and to bad transfer of the material to the substrate.
By various multimode adjustment of the laser resonator or using a
mask we formed the beam profile so that to get sharp and bright
lines ({\bf $\bigtriangledown$}, {\bf {\Large $\times$}},
{\bf $\bigcirc$}) or several  bright spots instead of the ordinary
({\LARGE $\bullet$}) filled-circle shape. As a result we got both the
high pulse energy density and the large size of the irradiated target
area without increasing the entire energy, so providing an oblong shape
of the plasma and an effective vapor
transfer. A reproducible growth of the high-quality
films was enabled (see examples in Table\,1 and Fig.\,2).

\begin{table}[h]
\caption{}

\begin{tabular}{|c|c|c|c|c|c|}

\hline
\,Beam profile\,  &       {\huge $_\bullet$}       &    {\huge $_\bullet$}  &      {\bf $\bigtriangledown$}   & {\bf {\Large $_\times$}} &   {\bf $\bigcirc$}    \\
\hline
Substrate         &    \,        YSZ         \,    & \,SrTiO$_\mathrm{3}$\, &    \,        YSZ         \,     & \,SrTiO$_\mathrm{3}$\,   & \,SrTiO$_\mathrm{3}$\,\\
\hline
                  &         77.0                   &       81.4             &        90.1                     &     91.3                 &         91.0          \\ 
$T_\mathrm{c}$, K &         86.1                   &       88.0             &        90.2                     &     90.8                 &         91.4          \\
                  &         82.7                   &       89.2             &        89.9                     &     91.0                 &         90.3          \\
                  &         87.1                   &       77.5             &        90.0                     &     90.7                 &         90.2          \\ 
\hline
\end{tabular}

\end{table}

\begin{figure}[hb]
\begin{center}\leavevmode
\includegraphics[width=1.0\linewidth]{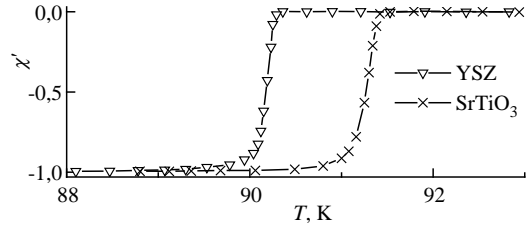}
\caption{ 
Alternating-field screening curves of YBCO films grown 
on YSZ ({\bf $\bigtriangledown$}-
profiled beam, $n$=400~Hz) and SrTiO$_\mathrm{3}$
({\bf \large $\times$}-profiling, $n$=380 Hz) substrates.
}
\label{fig2}
\end{center}
\end{figure}

\indent
Since our films grown on highly (100)-oriented substrates
had reduced $T_\mathrm{c}$ (85-87\,K), to all appearance
due to the terrace epitaxy disturbance\,\cite{magnet}, we became using
slanting cut substrates (all the data presented above belong to the films
grown on the substrates cut 2-4$^\mathrm{o}$ off the (100) plane). 
Stable growth of YBCO films with $T_\mathrm{c}$ about 
90~K or higher was observed on YSZ and SrTiO$_\mathrm{3}$
substrates cut with the slant from 2 to 20$^\mathrm{o}$.

\section{Conclusion}
In summary, due to the foregoing study of the on-axis
YAG:Nd$^\mathrm{3+}$ laser deposition of HTS films, 
several novelties (namely, the beam profiling, the velocity filtering
realized for the infrared exiting radiation, the considerably
non-stoichiometric targets, and the slanting cut substrates)
have been introduced and reproducible
growth of high-quality YBCO films has been reached.

\section{Acknowledgment}
The author is grateful to A.\,V.\,Varlashkin for the chopper control electronic
device assembling.

\end{document}